\documentstyle[aps,psfig]{revtex}
\input epsf.tex
\def \beq{\begin{equation}}
\def \eeq{\end{equation}}
\begin{document}
\baselineskip=24pt
\begin{center}
\bf{Dual variables for the $SU(2)$ lattice gauge theory at finite
temperature} \\
\vspace{1cm}

\rm{Srinath Cheluvaraja} \\
\it{Theoretical Physics Group\\
Tata Institute of Fundamental Research \\
Homi Bhabha Road,
Mumbai 400 005, India\\}
\end{center}
\vspace{1cm}

We study the three-dimensional $SU(2)$
lattice gauge theory at finite temperature using an observable which is
dual to the Wilson line. This observable
displays a behaviour which is the reverse of that seen for the Wilson line. 
It is non-zero 
in the confined phase and becomes zero in the deconfined phase.
At large
distances, it's
correlation function falls off exponentially in the deconfined phase and
remains non-zero in the confined phase.  The dual variable is
non-local and has a string attached to it which 
creates a $Z(2)$ interface in the system. It's correlation
function measures the string tension between oppositely
oriented $Z(2)$ domains. The construction of this variable can also
be made in the four-dimensional theory where it measures the surface tension
between oppositely oriented $Z(2)$ domains.

\vspace{5cm}
e-mail:srinath@theory.tifr.res.in
\newpage
\vspace{0.5cm}
Dual variables have played an important role in statistical mechanical
systems \cite{sav}. These variables display a behaviour which is the opposite
of that seen for the order parameters. They are non-zero in the disordered phase and
remain zero in the ordered phase.
Hence they are commonly referred to as disorder
variables. Unlike the order parameters which are local observables and
measure long range order in a statistical mechanical system, the dual
variables are non-local and are sensitive to disordering effects which often
arise as a consequence of topological excitations supported by a system -
like vortices, magnetic monopoles etc. Disorder variables for the $U(1)$ LGT
have been studied recently \cite{giac}. 
In this paper we study the finite
temperature properties of the three-dimensional $SU(2)$ lattice gauge theory
using an observable which is dual to the
Wilson line. We explain the sense in which this is dual
to the Wilson line and show that it's behaviour is the reverse of that
observed for the Wilson line. 
Unlike the Wilson line which creates a static quark
propagating in a heat bath, the dual variable creates a $Z(2)$ interface
in the system. The definition of this variable can also be extended
to the four-dimensional theory.

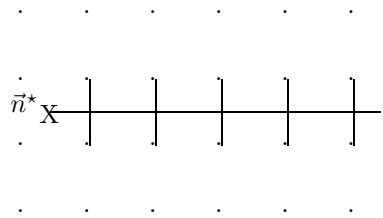
\begin{figure}
\begin{picture}(100,100)(-100,250)
\unitlength=5pt
\put(10,5){.}
\put(10,10){.}
\put(10,15){.}
\put(10,20){.}
\put(15,5){.}
\put(15,10){.}
\put(15,15){.}
\put(15,20){.}
\put(20,5){.}
\put(20,10){.}
\put(20,15){.}
\put(20,20){.}
\put(25,5){.}
\put(25,10){.}
\put(25,15){.}
\put(25,20){.}
\put(30,5){.}
\put(30,10){.}
\put(30,15){.}
\put(30,20){.}
\put(35,5){.}
\put(35,10){.}
\put(35,15){.}
\put(35,20){.}
\put(11.7,11.7){X}
\put(12.5,12.5){\line(1,0){25}}
\put(9.5,12.5){$\vec n^{\star}$}
\put(15.5,10){\line(0,1){5}}
\put(20.5,10){\line(0,1){5}}
\put(25.5,10){\line(0,1){5}}
\put(30.5,10){\line(0,1){5}}
\put(35.5,10){\line(0,1){5}}
\end{picture}
\vspace{4.5in}
\caption{Dual variable in the Ising model.}
\label{isidual}
\end{figure}

Before we consider the three-dimensional $SU(2)$ lattice gauge theory let
us briefly recall the construction of the dual variable for the 
two-dimensional Ising model \cite{kad}.
The variable dual to the spin variable $\sigma(\vec n)$ is
denoted by $\mu(\star \vec n)$ and is defined on the dual lattice. This
variable  which is shown in Fig.~\ref{isidual} has a string attached to it
which pierces the
bonds connecting the spin variables. The position of the string is not
fixed and it can be varied using a $Z(2)$ ($\sigma(\vec n)\rightarrow
-\sigma(\vec n)$) transformation. The average value of the dual variable
is defined as
\beq
<\mu(\star \vec n)>=\frac{Z(\tilde K)}{Z(K)}
\eeq
where $Z(K)$ and $Z(\tilde K)$ are the partition functions defined using
the coupling constants $K$ and $\tilde K$ respectively. 
The partition function for the Ising model is got from the Hamiltonian
\beq
H=-K\sum_{\vec n \vec n^{\prime}}\sigma (\vec n) \sigma (\vec n^{\prime}).
\eeq
The coupling
constant $\tilde K$ is defined as
\begin{eqnarray*}
\tilde K=-K&& \ \ on\  the\ bond\ pierced\ by\ the\ string \\ \nonumber
\tilde K=K&& \ \ elsewhere.
\end{eqnarray*}
The dual variable $\mu(\star \vec n)$ thus creates
an interface beginning from $\star \vec n$. It has the following
behaviour at high and low temperatures \cite{kad}
\begin{eqnarray*}
<\mu(\star \vec n)> \approx 1 \ for\ K\ small \\ \nonumber
<\mu(\star \vec n)> \approx 0 \ for\ K\ large.
\end{eqnarray*}
It is in this sense that the variable $\mu(\star \vec n)$ is dual to the
variable $\sigma(\vec n)$ which behaves as
\begin{eqnarray*}
<\sigma(\vec n)> \approx 0 \ for\ K\ small \\ \nonumber
<\sigma(\vec n)> \approx 1 \ for\ K\ large .
\end{eqnarray*}
The  spin and dual correlation functions satisfy the relation
\beq
<\mu(\star \vec n)\mu(\star \vec n^{\prime})>_{K>>1}=
<\sigma(\vec n)\sigma(\vec n^{\prime})>_{K<<1}.
\eeq
Using the $\sigma \rightarrow -\sigma$ transformation it can be shown that
the correlation function of the $\mu$'s is independent of the shape of the
string joining $\star \vec n$ and $\star \vec n^{\prime}$.
The variables $\sigma(\vec n)$ and $\mu(\star \vec n)$ satisfy the
algebra
\beq
\sigma(\vec n)\mu(\star \vec n)=\mu(\star \vec n)\sigma(\vec n)\exp (i\omega),
\eeq
where $\omega=0$ if the variable $\sigma$ does not lie on a bond pierced
by the string attached to $\mu(\star \vec n)$ and $\omega=\pi$ otherwise.

The above considerations generalize easily to the three-dimensional
$Z(2)$ gauge theory. The dual variables are again defined on the sites
of the dual lattice and the string attached to them will now pierce
plaquettes instead of bonds. Whenever a plaquette is pierced by a string
the coupling constant changes sign just as in the case of the Ising model. One
can similarly define correlation functions of these variables. Since the
three-dimensional $Z(2)$ gauge theory is dual to the the three-
dimensional Ising model, the correlation functions of these variables
will have a behaviour which is the reverse of the 
spin-spin correlation function in
the three-dimensional Ising model. For the case of the $SU(2)$ lattice
gauge theory which is our interest here,
the definition of these variables is more involved. However, since $Z(2)$
is a subgroup of $SU(2)$ one can define variables which are dual to the
$Z(2)$ degrees of freedom by following the same prescription as in the
three-dimensional $Z(2)$ gauge theory. The relevance and effectiveness
of these variables will depend on the role played by the $Z(2)$ degrees
of freedom in the $SU(2)$ lattice gauge theory. The role of the center
degrees of freedom in the $SU(2)$ lattice gauge theory
was also examined in \cite{mack}.

Since the finite temperature transition in $SU(N)$ lattice gauge theories
is governed by the center ($Z(N)$ for $SU(N)$) degrees of freedom\cite{suss}, we
expect these variables to be useful in studying this transition.
The  usual analysis of finite temperature lattice gauge theories is carried out
by studying the behaviour of the Wilson line which becomes
non-zero across the finite temperature transition \cite{suss}.
The non-zero value
of the Wilson line indicates deconfinement of static quarks.
The spatial degrees of freedom undergo no dramatic change across the
transition and only serve to produce short-range interactions between
the Wilson lines. Thus  one gets an effective theory of Wilson lines
in one lower dimension \cite{yaffe}.
The deconfinement transition can be monitored
by either measuring the expectation value of the Wilson line or by looking
at the behaviour of the Wilson line correlation function \cite{mac}.
In the confining phase, the correlation function is 
(for $|\vec n-\vec n^{\prime}|$ large)
\beq
<L(\vec n)L(\vec n^{\prime})> \approx \exp (-\sigma T|\vec n -\vec n^{\prime}|)
\eeq
while in the deconfining phase
\beq
<L(\vec n)L(\vec n^{\prime})> \approx constant .
\eeq

We define the variable $\mu(\star \vec n)$ on the dual lattice site $\star \vec
n$ as
\beq
\mu(\star \vec n)=\frac {Z(\tilde \beta)}{Z(\beta)}
\eeq

where $Z(\tilde \beta)$ is the partition function with couplings
$\tilde \beta$ which is defined as
\begin{eqnarray*}
\tilde \beta=-\beta&&\ on\ plaquettes\ pierced\ by\ the\ string \\ \nonumber
\tilde \beta=\beta&&\ elsewhere
\end{eqnarray*}
and the string runs in the spatial direction. Hence the plaquettes
pierced by the string are all space-time plaquettes.
The action for the $SU(2)$ LGT is chosen to be the Wilson action 
\cite{wils} which is
\beq
S=\frac{\beta}{2}\sum_{p}tr\ U(p).
\eeq
The variables $\mu(\star \vec n)$ and $L(\vec n)$ satisfy the
algebra
\beq
L(\vec n)\mu(\star \vec n)=\mu(\star \vec n)L(\vec n)\exp(i\omega)
\eeq
where $\omega=0$ if the plaquette pierced by the string attached to
$\mu (\star \vec n)$ is not touching any of the links belonging to
$L(\vec n)$ and $\omega=\pi$ if the plaquette makes contact with
any of the links of $L(\vec n)$. The variables $\mu(\star \vec n)$
and $L(\vec n)$ satisfy the same algebra as the $\sigma$ and $\mu$
variables in the Ising model. This is the same as the algebra of the
order and disorder variables in \cite{hoof}.
Note that this algebra is only
satisfied if the string is taken to be in the spatial direction.
The location of the string can again be changed by local $Z(2)$
transformations.
The correlation function of the dual variables is defined to be
\beq
<\mu (\star \vec x) \mu(\star \vec y)> = \frac{Z(\tilde \beta)}{Z(\beta)}
\label{disor}
\eeq
where $\tilde \beta =-\beta$ on all plaquettes pierced by the string
joining $\vec x$ and $\vec y$ and $\tilde \beta=\beta$ otherwise. It is again
easily seen that this quantity is independent of the shape of the string
which can always be varied by a $Z(2)$ ($U(n;\mu)\rightarrow -U(n;\mu)$)
transformation. The righthand side of Eq.~\ref{disor} can be expressed as
an average value
\beq
<\mu (\star \vec x) \mu(\star \vec y)> = <\exp(-\beta\sum_{p^{\prime}}tr\ U(p))>_{\beta}
\eeq
where the prime denotes that the summation is only over plaquettes
which are dual to the string joining $\vec x$ and $\vec y$,  and the average is taken
using the coupling $\beta$. Since the string is spatial, all the plaquettes
appearing in the sum are space-time plaquettes.

\begin{figure}
\begin{picture}(100,100)(-100,250)
\unitlength=5pt
\put(9,1){$\vec n$}
\put(19,1){$\vec n^{\prime}$}
\put(10,10){\line(0,1){50}}
\put(20,10){\line(0,1){50}}
\put(12,10){\line(0,1){10}}
\put(18,10){\line(0,1){10}}
\put(12,20){\line(1,0){6}}
\put(12,22){\line(1,0){6}}
\put(12,32){\line(1,0){6}}
\put(12,34){\line(1,0){6}}
\put(12,44){\line(1,0){6}}
\put(12,46){\line(1,0){6}}
\put(12,56){\line(1,0){6}}
\put(12,22){\line(0,1){10}}
\put(18,22){\line(0,1){10}}
\put(12,34){\line(0,1){10}}
\put(18,34){\line(0,1){10}}
\put(12,46){\line(0,1){10}}
\put(18,46){\line(0,1){10}}
\put(12,36){\line(1,0){6}}
\put(12,38){\line(1,0){6}}
\put(12,40){\line(1,0){6}}
\put(12,42){\line(1,0){6}}
\put(-10,32){$\tau \rightarrow$}
\end{picture}
\vspace{4.0in}
\caption{Strong coupling diagram in the finite temperature $SU(2)$ LGT
contributing to the effective Ising model.}
\label{strcoup}
\end{figure}
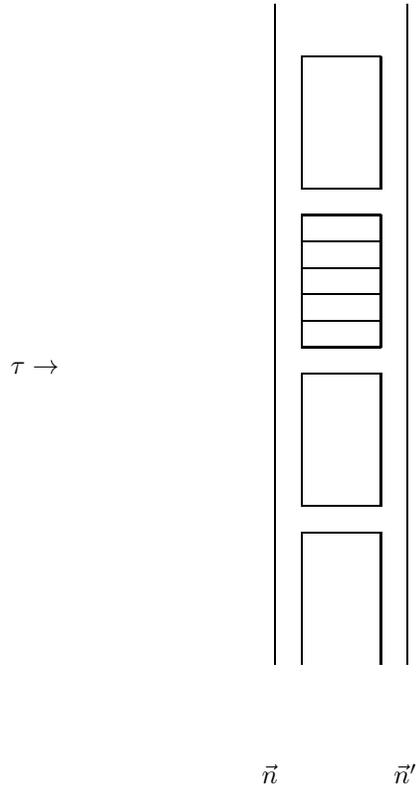

It is easy to show that the behaviour of this correlation function will
be the reverse of that of the Wilson line. To see this note that when the
spatial degrees of freedom are integrated out, we get an effective two-
dimensional model of Ising like spins with local interactions.
To leading order in strong coupling,
the effective action for the Wilson lines has the form
\beq
S_{eff}=2(\frac{\beta}{2})^{N_{\tau}}\sum_{\vec n \vec n^{\prime}}
J(\vec n-\vec n^{\prime})tr L(\vec n)
\ tr L(\vec n^{\prime}).
\label{effect}
\eeq
The term which gives this contribution is shown in Fig.~\ref{strcoup}.
When we calculate the correlation function in Eq.~\ref{disor} 
(where $\vec x$ and $\vec y$
are only separated in space) using
this approximation, one plaquette
occurring in this diagram will contribute with the opposite sign
(shown shaded in Figure.~\ref{strcoup}) and will cause the bond
between $\vec n$
and $\vec n^{\prime}$ to have a coupling with the opposite sign.
In Eq.~\ref{effect} $J(\vec n-\vec n^{\prime})$ contains the sign induced
on the bond.
This feature will persist for every diagram contributing to the effective
two-dimensional Ising model and it's effect will be to
create a disorder line
from $\vec x$ to $\vec y$.
Thus this correlation function will behave exactly like the disorder variable
in the two-dimensional Ising model and at large distances 
will fall off exponentially in the
ordered phase and will approach a constant value in the disordered phase.
We expect it to behave (for large $|\vec x-\vec y|$) as
\begin{eqnarray*}
<\mu(\vec x) \mu(\vec y)>\approx exp(-|\vec x-\vec y|/\xi)\ \ \  
\beta>\beta_{cr}
\\ \nonumber
<\mu(\vec x) \mu(\vec y)>\approx \mu^{2} \ \ \ \beta<\beta_{cr}
\end{eqnarray*}
Writing the above correlation function as
\beq
<\mu(\vec x) \mu(\vec y)>=\exp (-\beta_{\tau}(F(\vec x -\vec y))
\eeq
we can interpret $F$ as the free energy of an interface of length $|\vec x-
\vec y|$. The inverse temperature is denoted by $\beta_{\tau}$ to distinguish
it from the gauge theory coupling $\beta$.
In the ordered phase the interface energy increases
linearly with the length of the interface while in the disordered phase it is 
independent
of the length.
In the finite temperature system high temperature results in the
ordering of the Wilson lines and low temperature results in the disordering
of the Wilson lines. 
Therefore the dual variables will display ordering at low
temperatures and disordering at high temperatures.


A direct measurement of the dual variable results in large errors
because the dual variable is the exponential of a sum of plaquettes and
fluctuates greatly. We have directly measured the dual variable and the
correlation
function and found that they fall to zero at high temperatures and remain
non-zero at low temperatures. Since the measurement had large errors we 
prefer to use the method in \cite{giac2} where a similar problem was
encountered in the measurement of the disorder variable in the $U(1)$ LGT.
Instead of directly measuring the correlation function we measure
\beq
\rho(\vec x,\vec y)=-\frac{\partial ln<\mu>}{\partial \beta}.
\eeq
This quantity can be rewritten as
\beq
\rho(\vec x,\vec y)=-<\sum_{p\ne p^{\prime}}(1/2)tr\ U(p)>_{\tilde \beta}
+<\sum_{p =p^{\prime}}
(1/2)tr\ U(p)>_{\beta}
\eeq
where $p^{\prime}$ denotes the plaquettes which are dual to the string
joining $\vec x$ and $\vec y$. In our case this
quantity directly measures the free
energy of the $Z(2)$ interface between $\vec x$ and $\vec y$. Hence we expect
it to increase linearly with the interface length in the deconfining
phase and approach a constant value in the confining phase. Also this
variable is like any other statistical variable and is easier to
measure numerically. The variable $\rho$ can be used to directly measure
the interface string tension between oppositely oriented $Z(2)$ domains.
The behaviour of the quantity $\rho$ is
shown in Fig.~\ref{corr} and Fig.~\ref{corrb}. In the confined phase $\rho$ 
approaches a constant value at large distances while it increases linearly
with distance
in the deconfined phase. The slope of the straight line in Fig.~\ref{corr}
gives the interface string tension.
 The calculation of $\rho$ was made on a
$12**2\ 3$ lattice with 200000 iterations. The values of $\beta$ used were
$2.5$ in the confined phase and $5.5$ in the deconfined phase. The
deconfinement transition on the $N_{\tau}=3$ lattice occurs at $\beta=4.1$
\cite{hok}. The errors were estimated by blocking the data.

We would now like to point out a few applications of these dual variables.
The mass gap in the high temperature phase is determined
by studying the large distance behaviour of the Wilson line correlation
function. Since the Wilson line correlation function remains non-zero in the
deconfined phase the long distance part is subtracted out to get the
leading exponential. The dual variable correlation function already
displays an exponential fall off in the high temperature phase and 
provides us with another method of estimating the mass gap.
Also, since dual variables reverse the
roles of strong and weak coupling, they provide an alternate way of
looking at the system which may be convenient to address certain questions.
In this case they can be used to determine the string tension between
oppositely oriented $Z(2)$ domains in the $SU(2)$ gauge theory. The
surface tension between oppositely oriented $Z(2)$ domains in the four-
dimensional theory has been
calculated semi-classically in \cite{bhatt} .

The above construction of the dual variable 
can also be made in four dimensions. The only difference
is that in four dimensions the dual variables are defined on loops in the
dual lattice. The spatial string in three-dimensions is replaced by a
spatial surface which has the loops as it's the boundary. The dual variables
are functionals of the surface bounding the loops. The correlation
function of the dual variables is defined to be
\beq
<\mu (C,C^{\prime})>= <exp(-\beta \sum_{p^{\prime}}tr\ U(p))>
\eeq
where the summation is over all plaquettes which are dual to the
surface joining $C$ and $C^{\prime}$. Since the surface is purely
spatial the plaquettes contributing to the summation are all space-time
plaquettes. This correlation function will fall of exponentially
as the area of the surface joining $C$ and $C^{\prime}$ in the deconfined
phase and will approach a constant value in the confined phase.
A similar measurement of $\rho$ can be used to determine the surface tension
between oppositely oriented $Z(2)$ domains in the four-dimensional gauge theory.

\newpage
\newpage
\newpage
\setlength{\unitlength}{0.240900pt}
\ifx\plotpoint\undefined\newsavebox{\plotpoint}\fi
\sbox{\plotpoint}{\rule[-0.200pt]{0.400pt}{0.400pt}}%
\begin{figure}
\begin{picture}(1500,900)(0,700)
\font\gnuplot=cmr10 at 10pt
\gnuplot
\sbox{\plotpoint}{\rule[-0.200pt]{0.400pt}{0.400pt}}%
\put(177.0,158.0){\rule[-0.200pt]{4.818pt}{0.400pt}}
\put(155,158){\makebox(0,0)[r]{1.4}}
\put(1416.0,158.0){\rule[-0.200pt]{4.818pt}{0.400pt}}
\put(177.0,245.0){\rule[-0.200pt]{4.818pt}{0.400pt}}
\put(155,245){\makebox(0,0)[r]{1.6}}
\put(1416.0,245.0){\rule[-0.200pt]{4.818pt}{0.400pt}}
\put(177.0,332.0){\rule[-0.200pt]{4.818pt}{0.400pt}}
\put(155,332){\makebox(0,0)[r]{1.8}}
\put(1416.0,332.0){\rule[-0.200pt]{4.818pt}{0.400pt}}
\put(177.0,419.0){\rule[-0.200pt]{4.818pt}{0.400pt}}
\put(155,419){\makebox(0,0)[r]{2}}
\put(1416.0,419.0){\rule[-0.200pt]{4.818pt}{0.400pt}}
\put(177.0,507.0){\rule[-0.200pt]{4.818pt}{0.400pt}}
\put(155,507){\makebox(0,0)[r]{2.2}}
\put(1416.0,507.0){\rule[-0.200pt]{4.818pt}{0.400pt}}
\put(177.0,594.0){\rule[-0.200pt]{4.818pt}{0.400pt}}
\put(155,594){\makebox(0,0)[r]{2.4}}
\put(1416.0,594.0){\rule[-0.200pt]{4.818pt}{0.400pt}}
\put(177.0,681.0){\rule[-0.200pt]{4.818pt}{0.400pt}}
\put(155,681){\makebox(0,0)[r]{2.6}}
\put(1416.0,681.0){\rule[-0.200pt]{4.818pt}{0.400pt}}
\put(177.0,768.0){\rule[-0.200pt]{4.818pt}{0.400pt}}
\put(155,768){\makebox(0,0)[r]{2.8}}
\put(1416.0,768.0){\rule[-0.200pt]{4.818pt}{0.400pt}}
\put(177.0,855.0){\rule[-0.200pt]{4.818pt}{0.400pt}}
\put(155,855){\makebox(0,0)[r]{3}}
\put(1416.0,855.0){\rule[-0.200pt]{4.818pt}{0.400pt}}
\put(225.0,158.0){\rule[-0.200pt]{0.400pt}{4.818pt}}
\put(225,113){\makebox(0,0){1}}
\put(225.0,835.0){\rule[-0.200pt]{0.400pt}{4.818pt}}
\put(462.0,158.0){\rule[-0.200pt]{0.400pt}{4.818pt}}
\put(462,113){\makebox(0,0){2}}
\put(462.0,835.0){\rule[-0.200pt]{0.400pt}{4.818pt}}
\put(700.0,158.0){\rule[-0.200pt]{0.400pt}{4.818pt}}
\put(700,113){\makebox(0,0){3}}
\put(700.0,835.0){\rule[-0.200pt]{0.400pt}{4.818pt}}
\put(937.0,158.0){\rule[-0.200pt]{0.400pt}{4.818pt}}
\put(937,113){\makebox(0,0){4}}
\put(937.0,835.0){\rule[-0.200pt]{0.400pt}{4.818pt}}
\put(1175.0,158.0){\rule[-0.200pt]{0.400pt}{4.818pt}}
\put(1175,113){\makebox(0,0){5}}
\put(1175.0,835.0){\rule[-0.200pt]{0.400pt}{4.818pt}}
\put(1412.0,158.0){\rule[-0.200pt]{0.400pt}{4.818pt}}
\put(1412,113){\makebox(0,0){6}}
\put(1412.0,835.0){\rule[-0.200pt]{0.400pt}{4.818pt}}
\put(177.0,158.0){\rule[-0.200pt]{303.293pt}{0.400pt}}
\put(1436.0,158.0){\rule[-0.200pt]{0.400pt}{167.907pt}}
\put(177.0,855.0){\rule[-0.200pt]{303.293pt}{0.400pt}}
\put(45,506){\makebox(0,0){$\rho$}}
\put(806,23){\makebox(0,0){$x$}}
\put(177.0,158.0){\rule[-0.200pt]{0.400pt}{167.907pt}}
\put(225,212){\raisebox{-.8pt}{\makebox(0,0){$\Diamond$}}}
\put(462,473){\raisebox{-.8pt}{\makebox(0,0){$\Diamond$}}}
\put(700,547){\raisebox{-.8pt}{\makebox(0,0){$\Diamond$}}}
\put(937,648){\raisebox{-.8pt}{\makebox(0,0){$\Diamond$}}}
\put(1175,734){\raisebox{-.8pt}{\makebox(0,0){$\Diamond$}}}
\put(1412,819){\raisebox{-.8pt}{\makebox(0,0){$\Diamond$}}}
\put(225.0,174.0){\rule[-0.200pt]{0.400pt}{18.067pt}}
\put(215.0,174.0){\rule[-0.200pt]{4.818pt}{0.400pt}}
\put(215.0,249.0){\rule[-0.200pt]{4.818pt}{0.400pt}}
\put(462.0,447.0){\rule[-0.200pt]{0.400pt}{12.286pt}}
\put(452.0,447.0){\rule[-0.200pt]{4.818pt}{0.400pt}}
\put(452.0,498.0){\rule[-0.200pt]{4.818pt}{0.400pt}}
\put(700.0,512.0){\rule[-0.200pt]{0.400pt}{16.622pt}}
\put(690.0,512.0){\rule[-0.200pt]{4.818pt}{0.400pt}}
\put(690.0,581.0){\rule[-0.200pt]{4.818pt}{0.400pt}}
\put(937.0,617.0){\rule[-0.200pt]{0.400pt}{15.177pt}}
\put(927.0,617.0){\rule[-0.200pt]{4.818pt}{0.400pt}}
\put(927.0,680.0){\rule[-0.200pt]{4.818pt}{0.400pt}}
\put(1175.0,704.0){\rule[-0.200pt]{0.400pt}{14.213pt}}
\put(1165.0,704.0){\rule[-0.200pt]{4.818pt}{0.400pt}}
\put(1165.0,763.0){\rule[-0.200pt]{4.818pt}{0.400pt}}
\put(1412.0,791.0){\rule[-0.200pt]{0.400pt}{13.490pt}}
\put(1402.0,791.0){\rule[-0.200pt]{4.818pt}{0.400pt}}
\put(1402.0,847.0){\rule[-0.200pt]{4.818pt}{0.400pt}}
\end{picture}
\vspace{3.0in}
\caption{$\rho$ in the deconfining phase.}
\label{corr}
\end{figure}
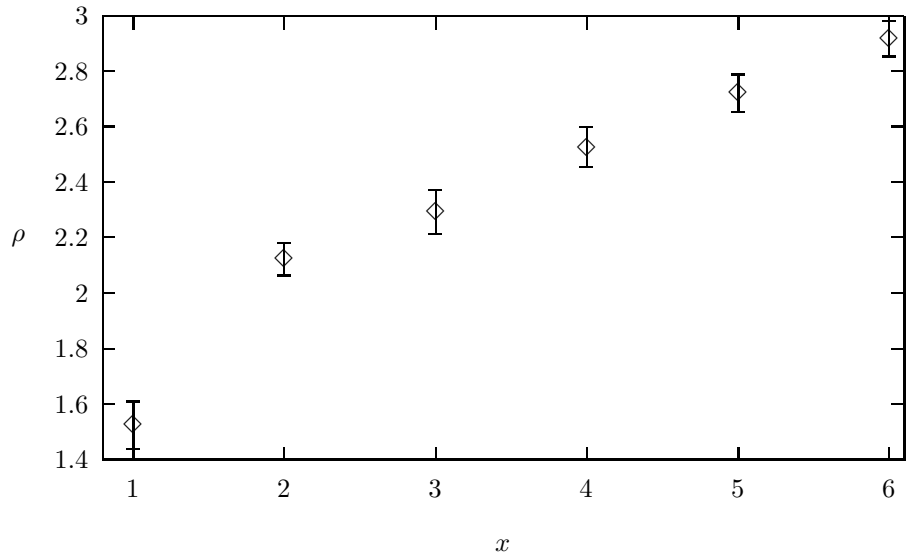
\newpage
\setlength{\unitlength}{0.240900pt}
\ifx\plotpoint\undefined\newsavebox{\plotpoint}\fi
\sbox{\plotpoint}{\rule[-0.200pt]{0.400pt}{0.400pt}}%
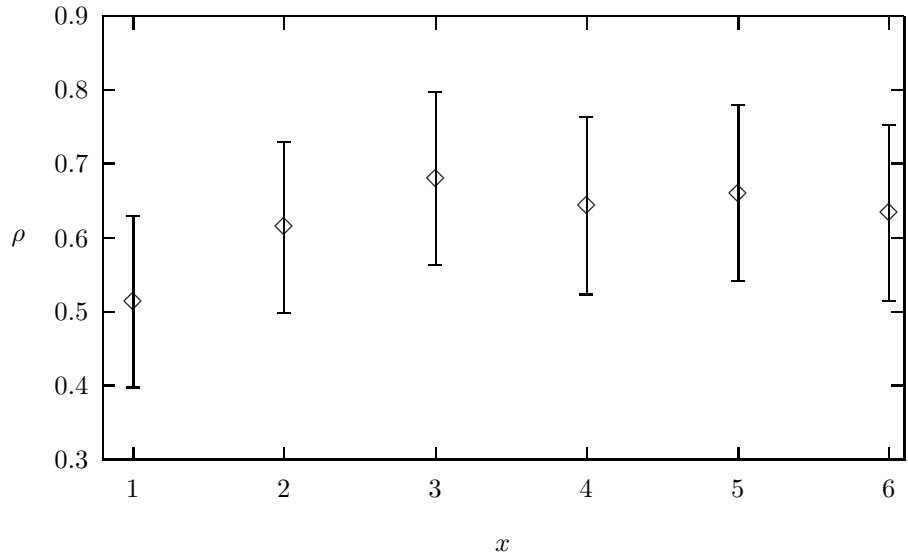
\begin{figure}
\begin{picture}(1500,900)(0,700)
\font\gnuplot=cmr10 at 10pt
\gnuplot
\sbox{\plotpoint}{\rule[-0.200pt]{0.400pt}{0.400pt}}%
\put(177.0,158.0){\rule[-0.200pt]{4.818pt}{0.400pt}}
\put(155,158){\makebox(0,0)[r]{0.3}}
\put(1416.0,158.0){\rule[-0.200pt]{4.818pt}{0.400pt}}
\put(177.0,274.0){\rule[-0.200pt]{4.818pt}{0.400pt}}
\put(155,274){\makebox(0,0)[r]{0.4}}
\put(1416.0,274.0){\rule[-0.200pt]{4.818pt}{0.400pt}}
\put(177.0,390.0){\rule[-0.200pt]{4.818pt}{0.400pt}}
\put(155,390){\makebox(0,0)[r]{0.5}}
\put(1416.0,390.0){\rule[-0.200pt]{4.818pt}{0.400pt}}
\put(177.0,506.0){\rule[-0.200pt]{4.818pt}{0.400pt}}
\put(155,506){\makebox(0,0)[r]{0.6}}
\put(1416.0,506.0){\rule[-0.200pt]{4.818pt}{0.400pt}}
\put(177.0,623.0){\rule[-0.200pt]{4.818pt}{0.400pt}}
\put(155,623){\makebox(0,0)[r]{0.7}}
\put(1416.0,623.0){\rule[-0.200pt]{4.818pt}{0.400pt}}
\put(177.0,739.0){\rule[-0.200pt]{4.818pt}{0.400pt}}
\put(155,739){\makebox(0,0)[r]{0.8}}
\put(1416.0,739.0){\rule[-0.200pt]{4.818pt}{0.400pt}}
\put(177.0,855.0){\rule[-0.200pt]{4.818pt}{0.400pt}}
\put(155,855){\makebox(0,0)[r]{0.9}}
\put(1416.0,855.0){\rule[-0.200pt]{4.818pt}{0.400pt}}
\put(225.0,158.0){\rule[-0.200pt]{0.400pt}{4.818pt}}
\put(225,113){\makebox(0,0){1}}
\put(225.0,835.0){\rule[-0.200pt]{0.400pt}{4.818pt}}
\put(462.0,158.0){\rule[-0.200pt]{0.400pt}{4.818pt}}
\put(462,113){\makebox(0,0){2}}
\put(462.0,835.0){\rule[-0.200pt]{0.400pt}{4.818pt}}
\put(700.0,158.0){\rule[-0.200pt]{0.400pt}{4.818pt}}
\put(700,113){\makebox(0,0){3}}
\put(700.0,835.0){\rule[-0.200pt]{0.400pt}{4.818pt}}
\put(937.0,158.0){\rule[-0.200pt]{0.400pt}{4.818pt}}
\put(937,113){\makebox(0,0){4}}
\put(937.0,835.0){\rule[-0.200pt]{0.400pt}{4.818pt}}
\put(1175.0,158.0){\rule[-0.200pt]{0.400pt}{4.818pt}}
\put(1175,113){\makebox(0,0){5}}
\put(1175.0,835.0){\rule[-0.200pt]{0.400pt}{4.818pt}}
\put(1412.0,158.0){\rule[-0.200pt]{0.400pt}{4.818pt}}
\put(1412,113){\makebox(0,0){6}}
\put(1412.0,835.0){\rule[-0.200pt]{0.400pt}{4.818pt}}
\put(177.0,158.0){\rule[-0.200pt]{303.293pt}{0.400pt}}
\put(1436.0,158.0){\rule[-0.200pt]{0.400pt}{167.907pt}}
\put(177.0,855.0){\rule[-0.200pt]{303.293pt}{0.400pt}}
\put(45,506){\makebox(0,0){$\rho$}}
\put(806,23){\makebox(0,0){$x$}}
\put(177.0,158.0){\rule[-0.200pt]{0.400pt}{167.907pt}}
\put(225,406){\raisebox{-.8pt}{\makebox(0,0){$\Diamond$}}}
\put(462,523){\raisebox{-.8pt}{\makebox(0,0){$\Diamond$}}}
\put(700,599){\raisebox{-.8pt}{\makebox(0,0){$\Diamond$}}}
\put(937,556){\raisebox{-.8pt}{\makebox(0,0){$\Diamond$}}}
\put(1175,576){\raisebox{-.8pt}{\makebox(0,0){$\Diamond$}}}
\put(1412,546){\raisebox{-.8pt}{\makebox(0,0){$\Diamond$}}}
\put(225.0,271.0){\rule[-0.200pt]{0.400pt}{65.043pt}}
\put(215.0,271.0){\rule[-0.200pt]{4.818pt}{0.400pt}}
\put(215.0,541.0){\rule[-0.200pt]{4.818pt}{0.400pt}}
\put(462.0,388.0){\rule[-0.200pt]{0.400pt}{64.802pt}}
\put(452.0,388.0){\rule[-0.200pt]{4.818pt}{0.400pt}}
\put(452.0,657.0){\rule[-0.200pt]{4.818pt}{0.400pt}}
\put(700.0,463.0){\rule[-0.200pt]{0.400pt}{65.525pt}}
\put(690.0,463.0){\rule[-0.200pt]{4.818pt}{0.400pt}}
\put(690.0,735.0){\rule[-0.200pt]{4.818pt}{0.400pt}}
\put(937.0,417.0){\rule[-0.200pt]{0.400pt}{67.211pt}}
\put(927.0,417.0){\rule[-0.200pt]{4.818pt}{0.400pt}}
\put(927.0,696.0){\rule[-0.200pt]{4.818pt}{0.400pt}}
\put(1175.0,438.0){\rule[-0.200pt]{0.400pt}{66.729pt}}
\put(1165.0,438.0){\rule[-0.200pt]{4.818pt}{0.400pt}}
\put(1165.0,715.0){\rule[-0.200pt]{4.818pt}{0.400pt}}
\put(1412.0,407.0){\rule[-0.200pt]{0.400pt}{66.729pt}}
\put(1402.0,407.0){\rule[-0.200pt]{4.818pt}{0.400pt}}
\put(1402.0,684.0){\rule[-0.200pt]{4.818pt}{0.400pt}}
\end{picture}
\vspace{3.0in}
\caption{$\rho$ in the confining phase.}
\label{corrb}
\end{figure}

\newpage
\begin{thebibliography}{99}
\bibitem{sav}{R.~Savit, Rev. Mod. Phys. {\bf 41}, 1 (1978). }
\bibitem{giac}{L.~Del Debbio, A.~D.~Giacomo and G.~Pafutti, Phys. Lett.
{\bf B349}, 513 (1995),Phys. Lett. {\bf B355},255 (1995).}
\bibitem{kad}{L.~Kadanoff and Ceva,Phys. Rev. {\bf B3},3918 (1971). }
\bibitem{mack}{G.~Mack and V.~Petkova, Ann. Phys. {\bf 125},  117 (1981).}
\bibitem{suss}{A.~Polyakov, Phys. Lett. {\bf 72B}, 477 (1978) ;
L.~Susskind, Phys. Rev. {\bf D20}, 2610 (1978).}
\bibitem{yaffe}{L.~Yaffe and B.~Svetitsky, Nucl. Phys.
{\bf B210}[FS6], 423 (1982).}
\bibitem{mac}{L.~Mclerran and B.~Svetitsky, Phys. Lett. {\bf 98B},
195 (1981) ; 
J.~Kuti, J.~Polonyi and K.~Szlachanyi, Phys. Lett. {\bf 98B},
199 (1981) ;
J.~Engels, F.~Karsch, H.~Satz and I.~Montvay, Nucl.
Phys.  {\bf B205}, 545 (1982).}
\bibitem{hoof}{G.~'t Hooft, Nucl. Phys. {\bf B138},1 (1978).}
\bibitem{wils}{K.~G.~Wilson, Phys. Rev. {\bf D10},2445 (1974).}
\bibitem{hok}{E.~D'Hoker, Nucl. Phys. {\bf B200}[FS4], 517 (1982).}
\bibitem{giac2}{A.~Giacomo and G.~Pafutti, hep-lat 9707003.}

\bibitem{bhatt}{T.~Bhattacharya et. al, Phys. Rev. Lett. {\bf 66},998 (1991);
T.~Bhattacharya et. al, Nucl. Phys. {\bf B383}, 497 (1992).}
\end {thebibliography}
\end{document}